\documentclass[a4paper,11pt]{article}
\pdfoutput=1

\usepackage{jinstpub}

\usepackage{textcomp}
\usepackage{float}
\usepackage{ifpdf}
\usepackage{caption}
\usepackage{subcaption}
\usepackage{amsmath}
\usepackage{soul}
\usepackage{comment}
\usepackage{lineno}

\def\sweep{{\sc sweep}}
\def\shift{{\sc shift}}
\def\locate{{\sc locate}}
\def\follow{{\sc follow}}
\def\tuning{{\sc tuning}}

\title{CONCERTO : Digital processing for finding and tuning LEKIDs}

\author[a,1]{ Julien Bounmy\note{Corresponding author}}
\author[a]{,  Christophe Hoarau}
\author[a]{,  Juan-Francisco Mac{\'{\i}}as-P{\'{e}}rez}
\author[c]{,  Alexandre Beelen}
\author[b]{,  Alain Beno\^it}
\author[a]{,  Olivier Bourrion}
\author[b]{,  Martino Calvo}
\author[a]{,  Andrea Catalano}
\author[c]{,  Alessandro Fasano}
\author[b]{,  Johannes Goupy}
\author[c]{,  Guilaine Lagache}
\author[a]{,  Julien Marpaud}
\author[b]{and  Alessandro Monfardini}

\affiliation[a] { CNRS, Univ. Grenoble Alpes, Grenoble INP\textsuperscript{$\dagger$}, LPSC-IN2P3 (Laboratoire de Physique Subatomique et de Cosmologie), 38000 Grenoble, France \\
$\dagger$ Institute of Engineering Univ. Grenoble Alpes}
\affiliation[b] { CNRS, Univ. Grenoble Alpes, Grenoble INP\textsuperscript{$\dagger$}, Institut Néel, 38000 Grenoble, France}
\affiliation[c] { Aix Marseille Univ., CNRS, CNES, LAM, Marseille, France}
\emailAdd{julien.bounmy@lpsc.in2p3.fr}

\abstract{We describe the on-line algorithms developed to probe Lumped Element Kinetic Inductance Detectors (LEKID) in this paper. LEKIDs are millimeter wavelength detectors for astronomy. LEKID arrays are currently operated in different instruments as: NIKA2 at the IRAM telescope in Spain, KISS at the Teide Observatory telescope in Tenerife, and CONCERTO at the APEX 12-meter telescope in Chile. LEKIDs are superconducting microwave resonators able to detect the incoming light at millimeter wavelengths and they are well adapted for frequency multiplexing (currently up to 360 pixels on a single microwave guide). Nevertheless, their use for astronomical observations requires specific readout and acquisition systems both to deal with the instrumental and multiplexing complexity, and to adapt to the observational requirements (e.g. fast sampling rate, background variations, on-line calibration, photometric accuracy, etc). This paper presents the different steps of treatment from identifying the resonance frequency of each LEKID to the continuous automatic control of drifting LEKID resonance frequencies induced by background variations.
}


\begin{document}
    \maketitle
    \flushbottom

\section{Introduction}

Lumped Element Kinetic Inductance Detectors (LEKID), which are exploited in millimeter astronomy, are superconducting microwave resonators and their resonance frequencies vary proportionally to the power of the absorbed incoming light.
So, the sky signal can be directly reconstructed from the observed shift in frequency of the detectors.
LEKIDs are developed in the form of arrays of hundreds to thousands pixels and can be used from few tens of GHz to THz wavelengths\citep{kids}.
For example, LEKID arrays are now operated in different instruments, such as: NIKA2 on the IRAM 30-meter telescope at Pico Veleta in Spain\citep{NIKA2gen}, the LEKIDs Interferometer Spectrum Survey (KISS) on the Q-U-I JOint TEnerife (QUIJOTE) 2.25-meter telescope at Teide Observatory in Tenerife \cite{Fasano_19JLTP_KISS}, and recently the CarbON CII line in the post-rEionization and ReionizaTiOn epoch project (CONCERTO) \cite{Lagache2020} at the Atacama Pathfinder EXperiment (APEX) 12-meter telescope located at 5\,100\,m above sea level on the Chajnantor plateau, Chile.   \\

Provided that LEKIDs can be naturally frequency multiplexed, the shifts in frequency for each detector in a single line can be measured simultaneously.
The measurement methodology relies on the injection of a probe signal, which is a frequency comb having each of its frequency adjusted to the LEKIDs resonators on the line, through the feed-line and analyze the returning signal to measure the variations of the transfer function of each LEKIDs.
These variations provide an accurate photometry \citep{fasano2021accurate}.
The reference frequency comb is obtained via the direct measurement of the line transfer function of the feed-line.

During standard astronomical observations the optical signal varies with time due to the influence  of the background sky emission and elevation angle changes.
As a consequence, the resonance frequency of the LEKIDs would be evolving during time and thus mitigation techniques are implemented to optimize working conditions and keep the frequency comb adjusted to the LEKIDs on the feed-line. At millimeter wavelengths for the NIKA \cite{Monfardini_2011} and NIKA2 \cite{NIKA2gen} photometric cameras specific treatment was implemented \cite{Catalano2014}.
At optical wavelength for the MKID Exoplanet camera \cite{MKID_SUBARU}, with different specifications and needs, algorithms using deep learning techniques are used \cite{MKID_TUNING}.
In the case of CONCERTO, we have developed the \sweep, \locate, \follow\ and \tuning\ procedures, which are presented in this paper.

First, at the beginning of each observation campaign, which lasts 10 to 15 days, the instrument is pointed to the sky and the individual LEKIDs resonance frequencies are determined with a dedicated large frequency \sweep\ for a starting reference point. The frequency comb obtained is then used as the reference comb for all subsequent fast adjustment operations.
Then, at the beginning of each set of observations (typically once per day) the instrument is pointed to the sky and a reduced frequency scan is performed in order to determine a global comb frequency adjustment with respect to the reference. This is the \locate\ procedure. 
Then the instrument goes in \follow\ procedure, where this global frequency adjustment is recomputed several times per second in order to compensate for changing background during telescope movements.
Both the \locate\ and \follow\ procedures are based on the \shift\ algorithm only they use its results in different manners at different time.
Finally, as soon as the target is pointed and prior to integration, a more precise frequency adjustment procedure is executed to determine individual frequency shift to apply.
This \tuning\ procedure takes about 30 seconds to complete.

In this paper, we discuss in details the above procedures by focusing on the case of the CONCERTO instrument. 
In Section~\ref{sec:lekids}, we give a brief description of the LEKIDs and arrays. Section~\ref{sec:readout} introduces the electronic readout system. 
In Section~\ref{sec:sweep}, we present the \sweep\ procedure that allows us to define a a reference comb saved in a file corresponding to LEKID resonance frequencies at a starting reference point.
We then present the modulation points principle, which is a system that we have developed.
In the last part, we describe the \shift\ and \tuning\ methods that are online algorithms that we use to cope with the background variation.

\section{Description of LEKID}
\label{sec:lekids}
\begin{figure}[hbtp] \centering
    \includegraphics[ width=0.29\textwidth]{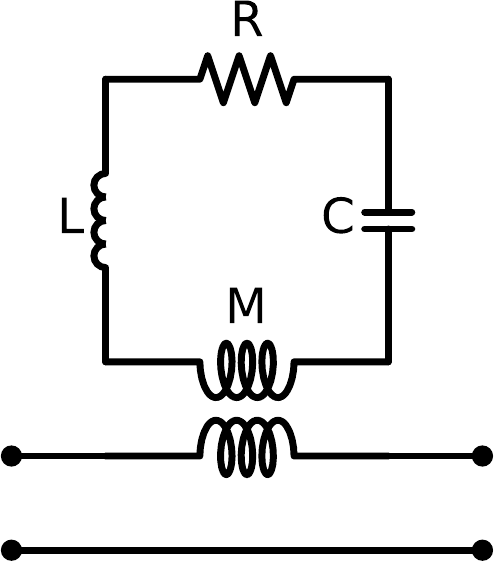}
    \includegraphics[ width=0.37\textwidth]{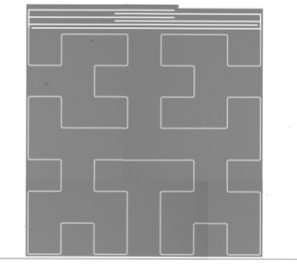}
\caption{ Schematic (l.h.s) and picture (r.h.s) of a LEKID. \label{figLEKID}} 
\end{figure}

The CONCERTO detectors consist of matrixes of LEKIDs placed in series on a microwave guide (microstrip feedline). Each LEKID pixel in the array is a microwave lumped element resonator composed of an inductor and an interdigitated capacitor, which is coupled to microstrip feedline (see figure~\ref{figLEKID}). Each LEKID fits on a layout of around one to ten square millimeters,  matched to the targeted detection wavelength.
The principle of detection is based on the fact that the absorption of a photon with an energy equal or greater to the band-gap of the superconductor breaks Cooper pairs. This affects the conduction, i.e. a fraction of the kinetic inductance turns into resistance, which in turn modify the resonance frequency of the detector, which is given by:
\begin{equation}
   f_0 = \frac{1}{2 \pi \sqrt{L C}}
    \label{eqF0}
\end{equation}
For LEKIDs, the transmission coefficient or scattering parameter $S_{21}$, can be expressed as: 
\begin{equation} \label{eqS21Res}
   S_{21} = \frac{ 1 + {\rm j} \omega R C + ({\rm j} \omega)^2 L C }{ 1 + {\rm j} \omega C \left( R+ \frac{(\omega M)^2}{2 Z_0} \right) + ({\rm j} \omega)^2 L C}
\end{equation}
with $\omega$ the pulsation and $Z_0$ the ports characteristic impedance.
This equation can be written in canonical shape 
\begin{equation} \label{eqFTRes}
   S_{21} = \frac{1 +{\rm j} \frac{1}{Q_{I}} \frac{\omega}{\omega_0} + \left({\rm j} \frac{\omega}{\omega_0}\right)^2}{1 +{\rm j}\frac{\omega}{\omega_0} \left(\frac{1}{Q_{I}}+\frac{1}{Q_{C}}\right) + \left({\rm j} \frac{\omega}{\omega_0}\right)^2}
\end{equation}
with  $\omega_0=\frac{1}{\sqrt{L C}}$, the resonance pulsation, $Q_{I} =\frac{1}{R} \sqrt{ \frac{L}{C}}$, the intrinsic quality factor and $Q_{C} =\frac{2 Z_0}{(\omega M)^2} \sqrt{ \frac{L}{C}}$, the quality factor of the coupling to the transmission line. \\

\begin{figure}[hbtp] \centering
    \includegraphics[ width=0.43\textwidth]{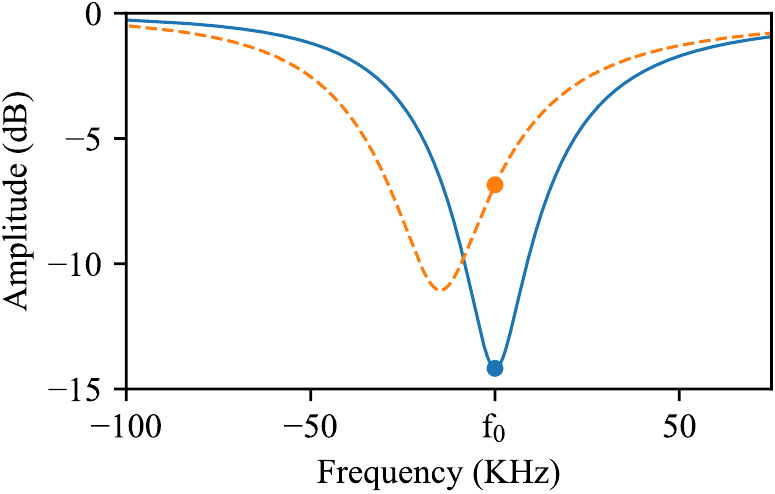}
    \includegraphics[ width=0.3\textwidth]{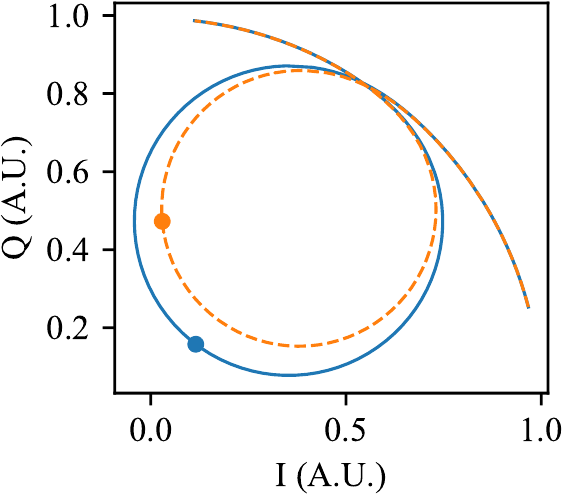}
\caption{ An example of a LEKID displayed around its resonance frequency in the scatter parameter $S_{21}$ in IQ plan. In blue, the LEKID is in a lower background condition; in orange dashed with a higher background.} \label{figOneTone}
\end{figure}

The resonance frequency is adjusted/tuned at the design stage by changing the capacitance.
This is achieved by modifying the length of the interdigitated capacitor. To probe a resonator, a sinusoidal wave is generated and injected into the feedline, at the resonance's frequency.
The output and modified signal is analyzed with as many Digital Down Converters (DDC) as LEKIDs on the instrumented line.
Each of these DDC produces In-phase (I) and Quadrature (Q) component of each modified signal, therefore, it is possible to visualize the change of a resonance, simulating a detection, in the IQ plan, as shown in figure~\ref{figOneTone}.

\section{Electronic readout and modulation}
\label{sec:readout}
\begin{figure}[hbtp] \centering
    \includegraphics[ width=0.99\textwidth]{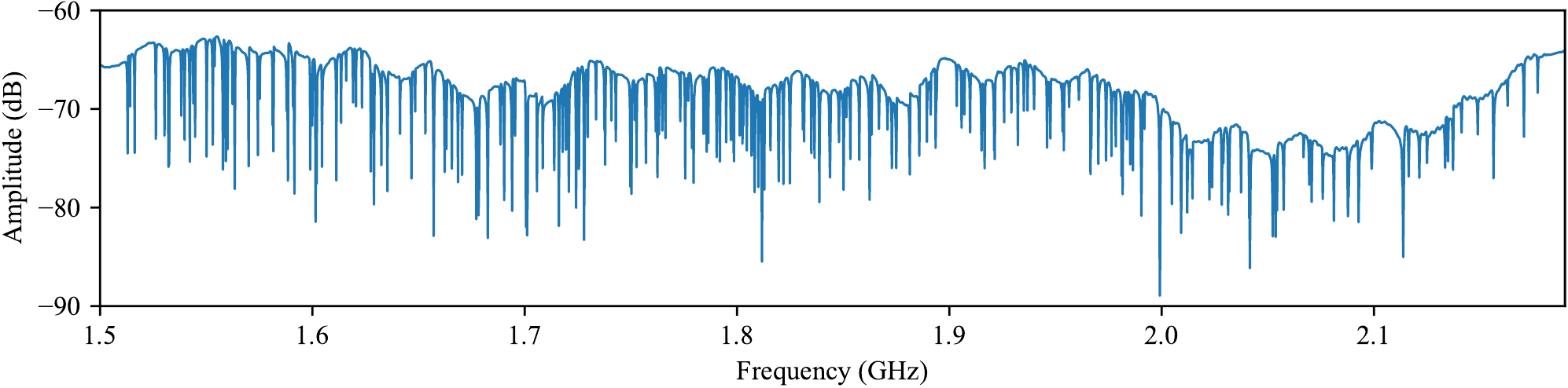}
\caption {Transmission coefficient of a typical feed-line, i. e. $\rm S_{21}$. Each dip in the figure corresponds to the response of a LEKID.} \label{figAllTones}
\end{figure}

The CONCERTO camera module consists of two focal planes separated by a polarizer.
Each of them is sampled by a LEKID array of 2152 resonators subdivided in six feed-lines.
Each feed-line is filled with around 360 LEKID resonators distributed on a 1\,GHz bandwidth located between 1.5 and 2.5\,GHz (see  figure~\ref{figAllTones}).

An electronic readout board, named KID\_READOUT, is dedicated for each feed-line.
It generates an IQ frequency comb composed of up to 400 tones, i. e. sinusoidal waveform, on a 1\,GHz band.
The electronic board can adjust each tone's frequency so each one of them can probe a LEKID resonator even if its resonance frequency drifts \cite{2016JInst..1111101B,concertoElec,2012JInst...7.7014B}. 

The comb signal is up-converted to the LEKID's resonance frequency band (1.5--2.5\,GHz) by a Local Oscillator (LO) generated from a programmable synthesizer.
Consequently, each tone's frequency is shifted with the same $F_{LO}$ =1.5GHz.

Additionally, we also modulate the LO's frequency with a digital signal [REF ELEC] so we can probe two more points around LEKIDs at +/-$F_{mod}$. Such technique was introduced in the NIKA2 experiment where only two modulation points were used, it permits to deduce the photometry from the frequency shift as explained in \cite{improvedMMphoto_Calvo}. The third unmodulated point was added in the KISS experiment \cite{fasano2021accurate}, this is the method on which the CONCERTO's three-point-modulation system is based. In practice, it allows us to sample three points ($p_1$, $p_2$ and $p_3$) around every LEKIDs resonance frequency.

\begin{figure}[hbtp] \centering
    \hspace*{-0cm}
    \includegraphics[ width=1.00\textwidth]{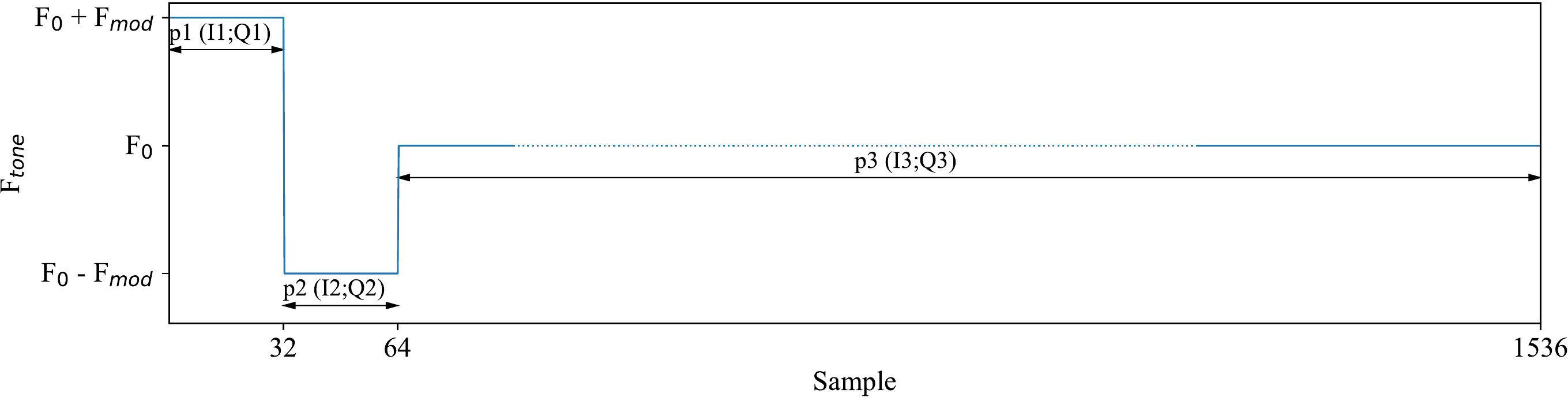}
\caption {Overview of the injected frequency modulation as a function the sample number. $F_{tone}$ is modulated at the beginning of each  data block (composed of 1536 samples). The sampling rate is 3.816 kHz, giving a 2.48 Hz data block rate.  \label{figModulation}}
\end{figure}

For CONCERTO, the three-point-modulation is operated in three parts along the 1536 samples collected to form a data block as shown in figure~\ref{figModulation}. It can be expressed as:
\begin{itemize}
\item $p_1$ ($I_1$;$Q_1$) : 32 positive modulation points at $F_{tone}$ = $F_{0}$ + $F_{mod}$
\item $p_2$ ($I_2$;$Q_2$) : 32 negative modulation points at $F_{tone}$ = $F_{0}$ - $F_{mod}$
\item $p_3$ ($I_3$;$Q_3$) : 1472 unmodulated observation points at $F_{tone}$ = $F_{0}$
\end{itemize}

\begin{figure}[hbtp] \centering
    \includegraphics[ width=0.32\textwidth]{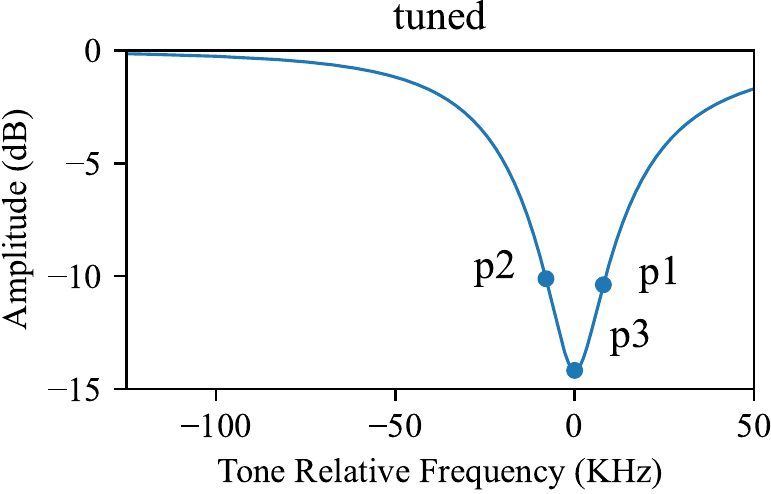}
    \includegraphics[ width=0.32\textwidth]{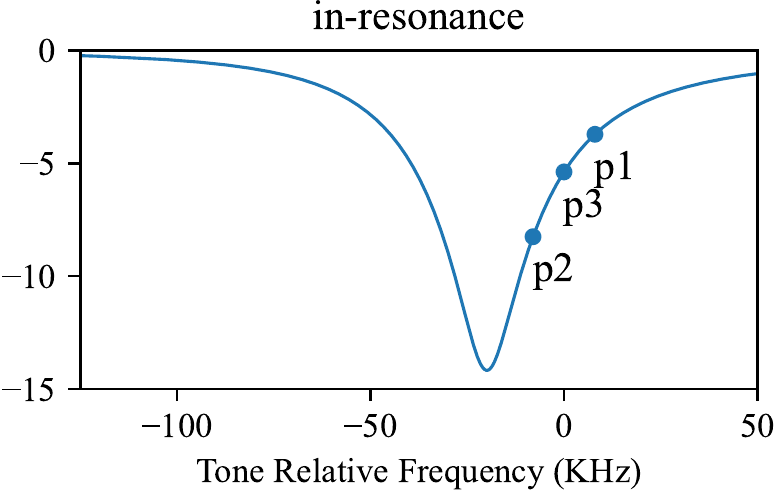} 
    \includegraphics[ width=0.32\textwidth]{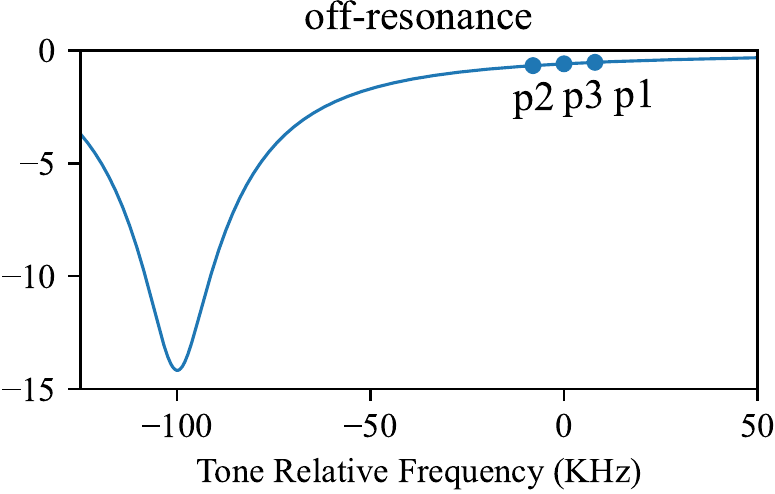} 
\caption{The LEKID response to the three-point-modulation is shown for three cases. {\it tuned} when $F_{0}$ = $F_{res}$ (l.h.s), {\it in-resonance} when $| F_{0}-F_{res} | < \Delta F_{res}$ and  {\it off-resonance} (r.h.s)  when $|F_{0}-F_{res} | >> \Delta F_{res}$.  \label{figToneMod}}
\end{figure}

For CONCERTO $F_{mod}$ is generally set at about 10\,kHz.
The modulation scheme can be used both for calibration purposes (see~\cite{fasano2021accurate}) and to ensure that the LEKID response is correctly sampled at the resonance frequency.
In this regard and for the sake of simplicity, we consider here three main cases as shown in figure~\ref{figToneMod}.
A perfectly sampled LEKID resonance frequency when $F_{0} = F_{res}$, called {\it tuned} hereafter.
The {\it in-resonance} case when $| F_{0}-F_{res} | < \Delta F_{res}$ where $\Delta F_{res}$ corresponds to the width of the LEKID response.
And finally, the {\it off-resonance} case when $| F_{0}-F_{res} | >> \Delta F_{res}$. 

The main goal of the CONCERTO acquisition system is to operate the LEKIDs as close as possible to the {\it tuned} case.
The three-point-modulation is used to achieve this goal by estimating $| F_{0}-F_{res} |$.
Indeed, as shown in figure~\ref{figToneMod} for the {\it tuned} and {\it in-resonance} cases, the $p_1$, $p_2$ and $p_3$ provide an information about the shape of the resonance and its frequency shift.
Unfortunately, in the {\it off-resonance} case, these points are smeared by the measurement noise and useless to infer the resonance position, hence making difficult the correction estimation.

For all the techniques discussed in this paper, it is important to note that each tone's excitation frequency is digitally generated by the readout electronics.
Therefore the frequency resolution is quantified and in our case limited at 3.816\,kHz.
This yields a maximum uncertainty of half that frequency to reconstruct $F_{res}$.

\section{Determining the LEKID resonance frequencies via a \sweep\ }
\label{sec:sweep}

The first step for operating the LEKIDs is to identify their resonance frequencies via a \sweep\ , which is a reconstruction of the transmission coefficient, i. e. scattering parameter $S_{21}$. 
To optimize the \sweep\ , 400 tone frequencies are placed equidistantly over the 1\,GHz bandwidth and the whole frequency comb is shifted on a range of 2.5\,MHz, which is sufficient to scan the whole 1\,GHz bandwidth.
This method is faster than using a single tone to sweep the full 1\,GHz bandwidth.
Indeed, the successive frequency shifts are performed by gradually changing the LO frequency, therefore reducing the frequency range to cover allow both a finer and faster sweep.

The resulting data consist of a set of I and Q vectors for each tone, which can be associated to a given frequency:
$F_{tone}(t) = F_{tone}(0) + \Delta F_{LO} (t) $, where $\Delta F_{LO} (t)$ is the LO frequency shift.
The I and Q vectors can be transformed in amplitude A,  and phase $\Phi$ such that $(I,Q) = A \times (\cos{(\Phi)}, \sin{(\Phi)})$.

\begin{figure}[hbtp] \centering
\begin{subfigure}[b]{0.32\textwidth} \centering
    \includegraphics[width=\textwidth]{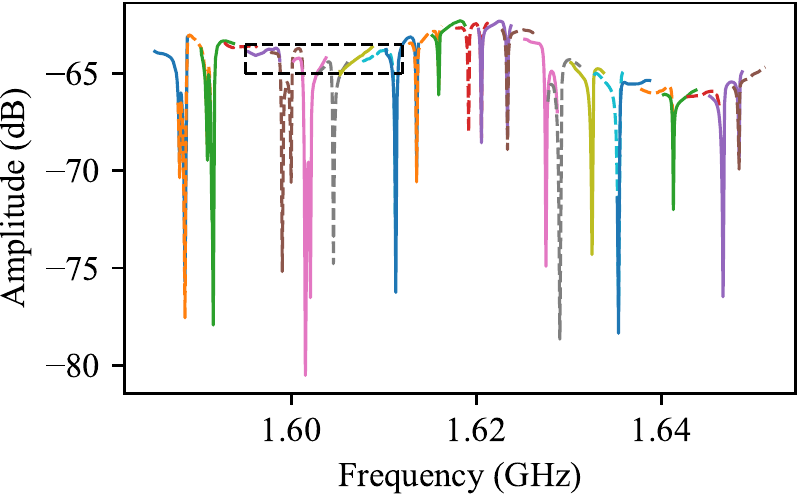}
    \includegraphics[width=\textwidth]{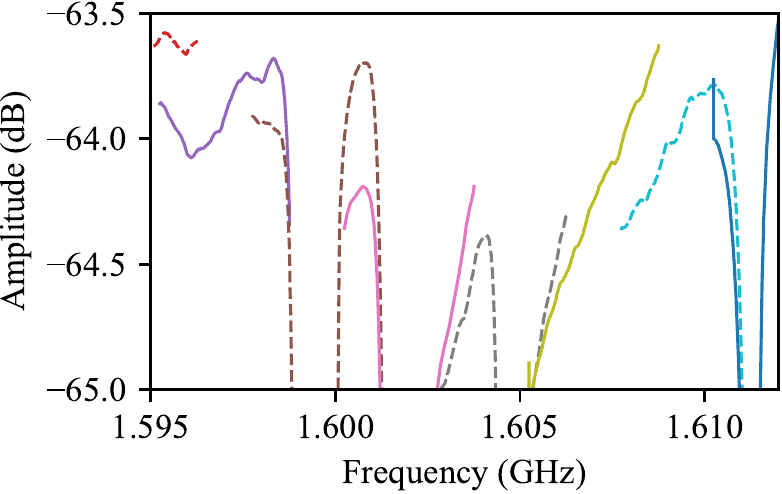}
    \caption{Raw data per tone} \label{figTonesRAW}
\end{subfigure}
\begin{subfigure}[b]{0.32\textwidth} \centering
    \includegraphics[width=\textwidth]{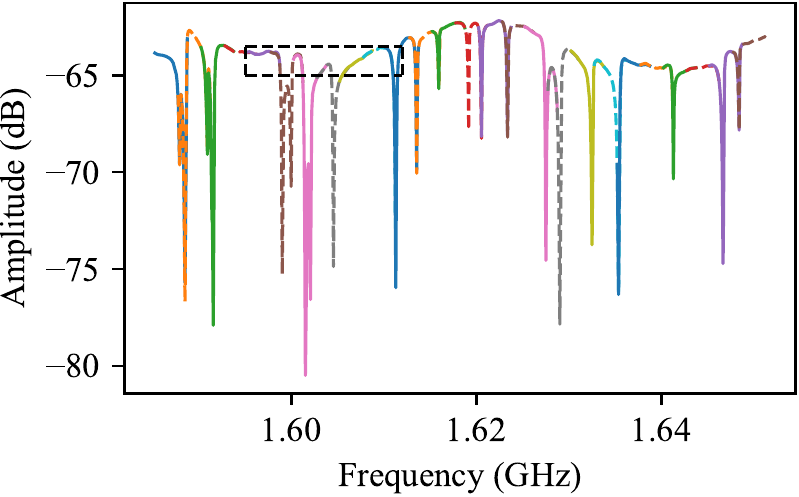} 
    \includegraphics[width=\textwidth]{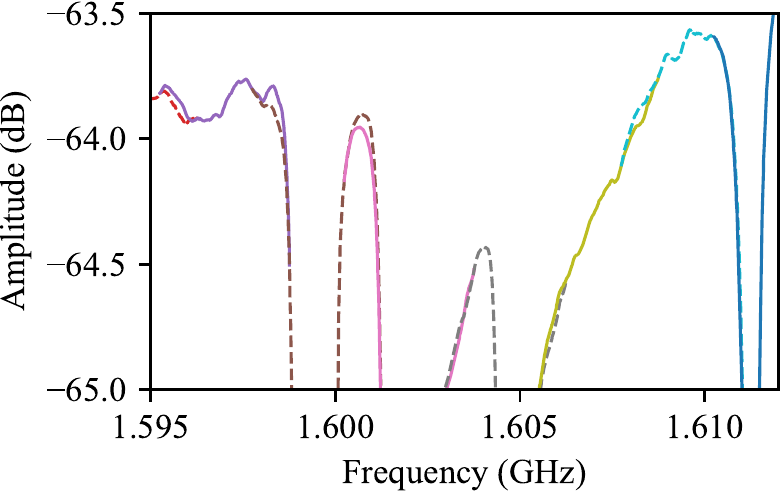}
    \caption{Offset corrected data per tone} \label{figTonesCOR}
\end{subfigure}
\begin{subfigure}[b]{0.32\textwidth} \centering
    \includegraphics[width=\textwidth]{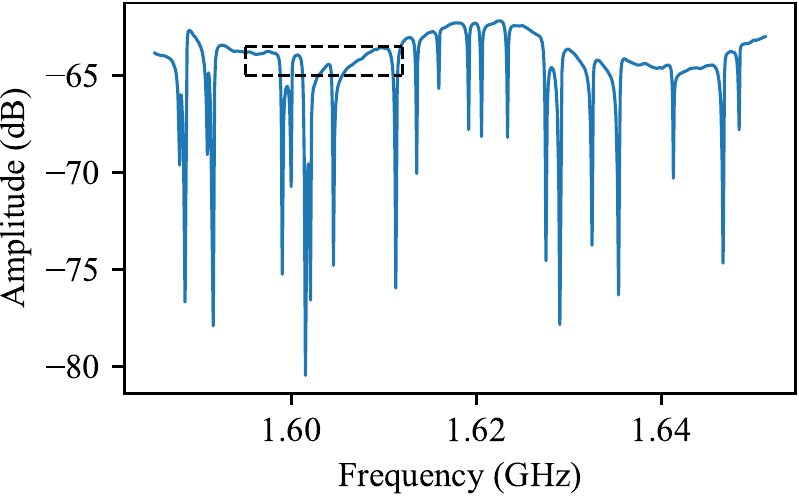} 
    \includegraphics[width=\textwidth]{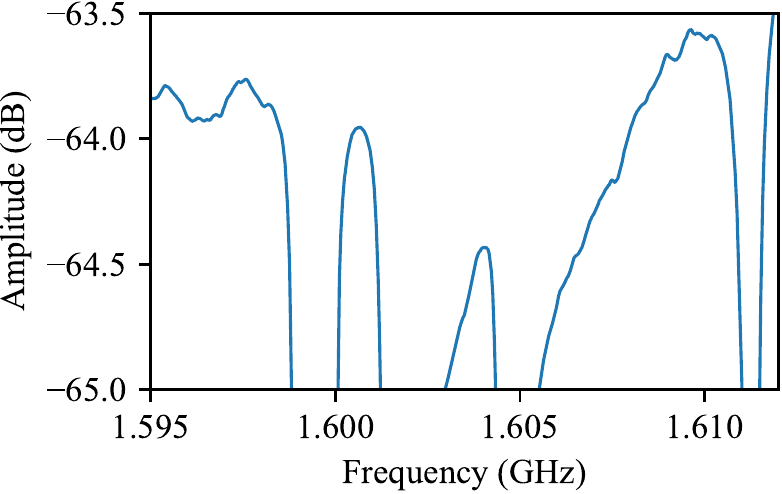}
    \caption{Concatenated data} \label{figTonesConcat}
\end{subfigure}
\caption{Top: \sweep\ construction steps from raw data (one vector per tone) to concatenated frequency data. Bottom: Zoom of the top figures within the black dashed rectangle.} \label{figTones}
\end{figure}

We show in figure~\ref{figTonesRAW} the raw amplitude as a function of frequency for several tones, each tone is represented with a different color.
Unfortunately, as a consequence of the optimized \sweep, there are discontinuities between tones and these need to be corrected before any further processing. 
The origin of these discontinuities are due to the small variations of the base band electronic comb.
Hence, we correct for offsets in phase and amplitude to smooth out these discontinuities, see figure~\ref{figTonesCOR}).
Finally, the corrected raw data of each tones have to be concatenated to construct a single vector as function of frequency which is the transmission coefficient, see figure \ref{figTonesConcat}.

\begin{figure}[hbtp] \centering
    \includegraphics[width=0.32\textwidth]{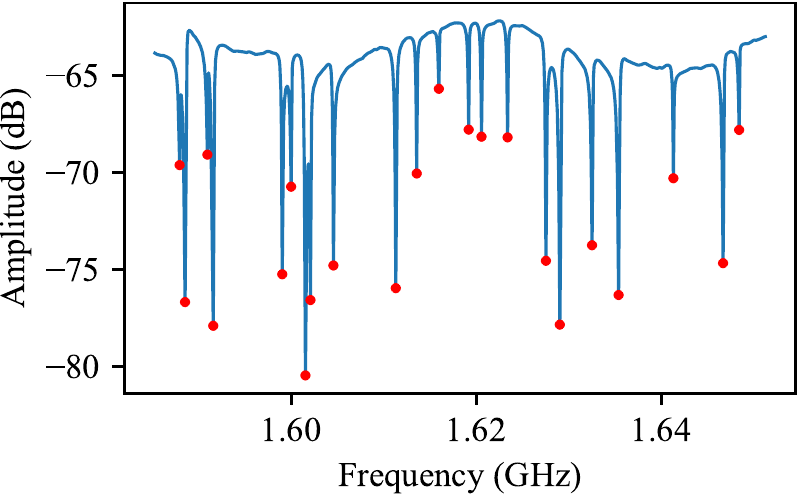}
    \includegraphics[width=0.32\textwidth]{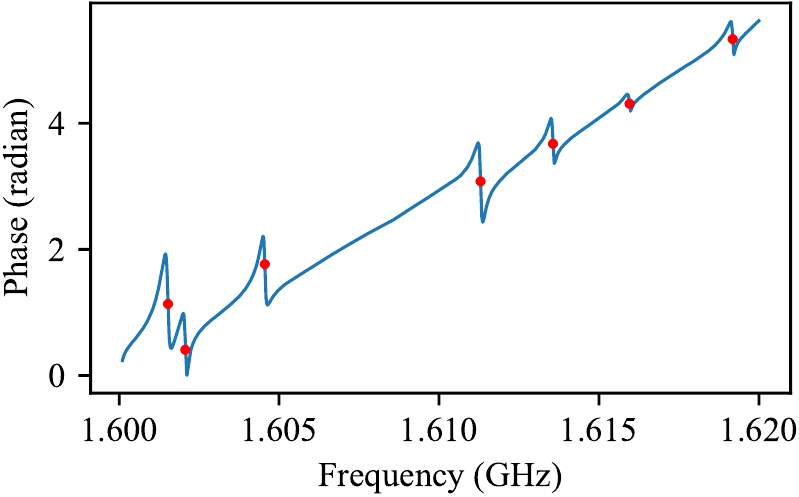}
    \includegraphics[width=0.32\textwidth]{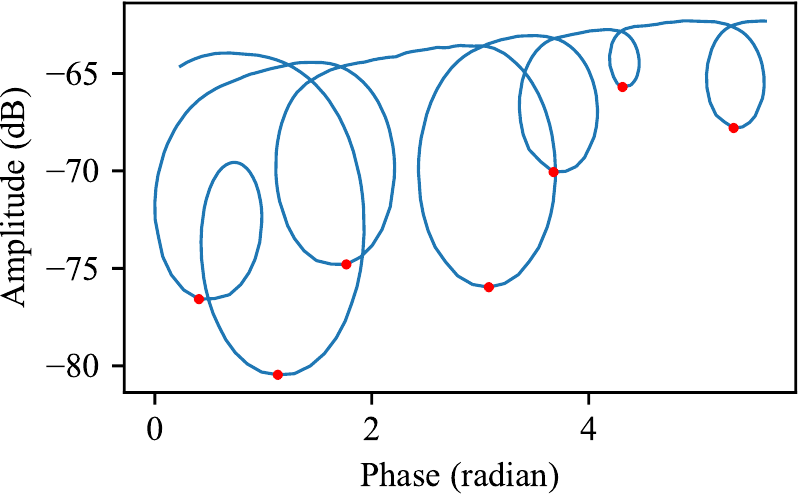}
\caption{Data for a \sweep\ in amplitude (left) and phase (middle) as a function of frequency, and amplitude as a function of phase (right). } \label{figSweep}
\end{figure}

The concatenated phase and amplitude data is used to extract the resonance frequency of each detector.
For that,we consider two main effects: 1) at the resonance frequency, the amplitude of the LEKID response goes through a minimum, and 2) the slope of the phase through a maximum. 
This is  illustrated in figure~\ref{figSweep} where the resonance frequency for each LEKID is represented by a red dot.
This can also be expressed as a zero on the derivative of the amplitude through frequency and a maximum on the 
derivative of the phase through frequency.

The algorithm to find the LEKID resonance frequency is to compute a first order forward difference on the phase as function of frequency.
Then, to increase reliability and accelerate the process, a sliding average filtering is applied on the derivative data to reduce noise.
Afterward, the smoothed data are used to search for the maximums in the phase variation, that give a first rough localisation of the resonance frequencies. 
Finally, a fine localisation is performed by searching for for local minimums in the amplitude data.

As a result of this procedure, we can generate a configuration file holding the LEKID resonance frequencies for this starting reference point.
During operations, at start up, background conditions are expected to differ with respect to those of the \sweep\ and thus the LEKIDs resonance frequencies are also expected to shift. The overall time performance in current software on CONCERTO is about fifteen minutes. 
Nevertheless, the relative distance between the resonances stays globally the same.
So, instead of redoing a full \sweep, which is a very slow process, we still use the saved configuration file and  correct for background variations using the \shift\ procedure described in the following section.

\section{Adapting to LEKIDs resonance frequency via \shift}

Large background variations leading to an overall shift of the LEKIDs resonance frequencies are expected in various situations.
This is for example the case at the beginning of the observing session after uploading the \sweep\  reference tones as discussed above.
Similarly, we expect significant background changes between two consecutive scans because either the source moves in elevation or we change target.
In order to place the tones as close as possible to the resonance frequency of the LEKIDs we shift the comb globally in frequency.

\begin{figure}[htb]
  \begin{subfigure}[b]{0.3\linewidth}
    \centering
    \includegraphics[width=\linewidth]{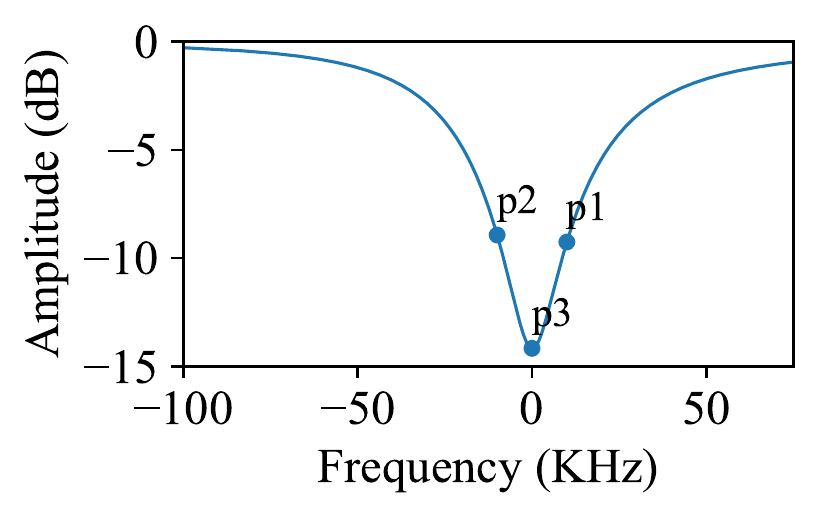}
    \vspace*{\baselineskip} 
    \includegraphics[width=0.68\linewidth]{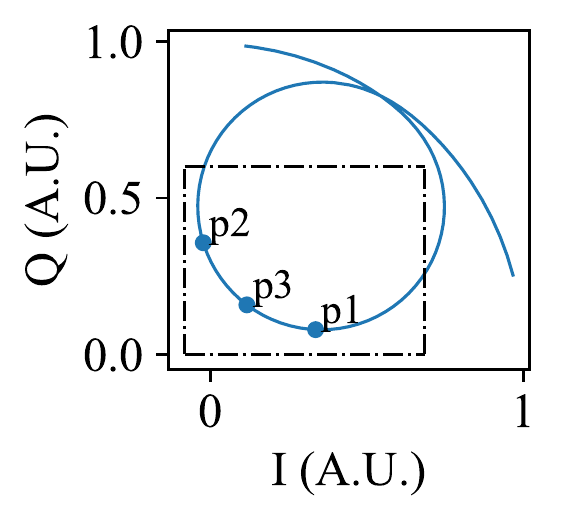}    
  \end{subfigure}
  \begin{subfigure}[b]{0.7\linewidth}
    \centering
    \includegraphics[width=0.89\linewidth]{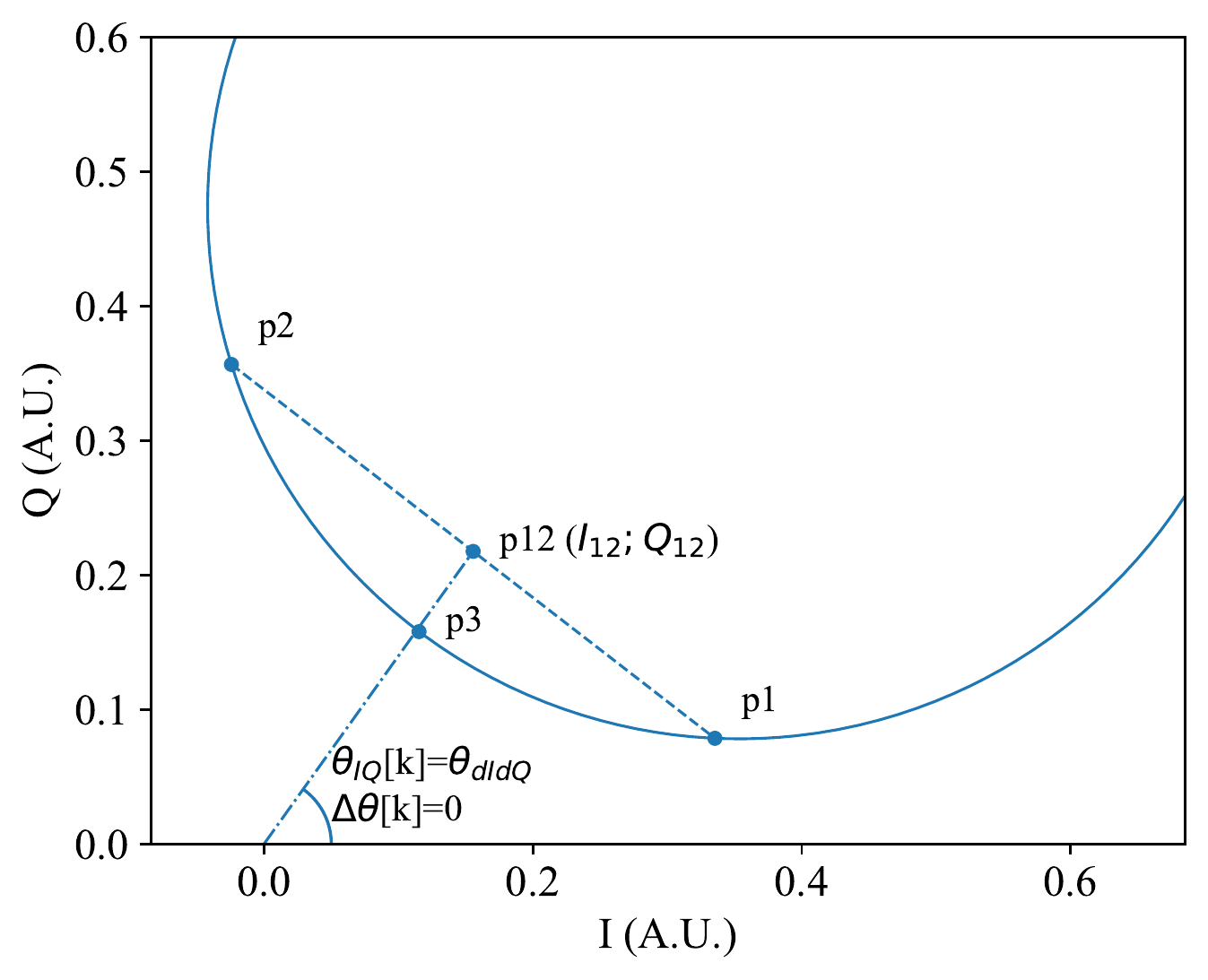}
  \end{subfigure}%
  \caption{An example of a LEKID correctly probed, $p_3$ is at the bottom of the resonance (F$_{tone}$ = F$_{res}$). This is the case when $\Delta\theta$[k] = 0.}
\label{figZoomRON}
\end{figure}

\begin{figure}[htb]
  \begin{subfigure}[b]{0.3\linewidth}
    \centering
    \includegraphics[width=\linewidth]{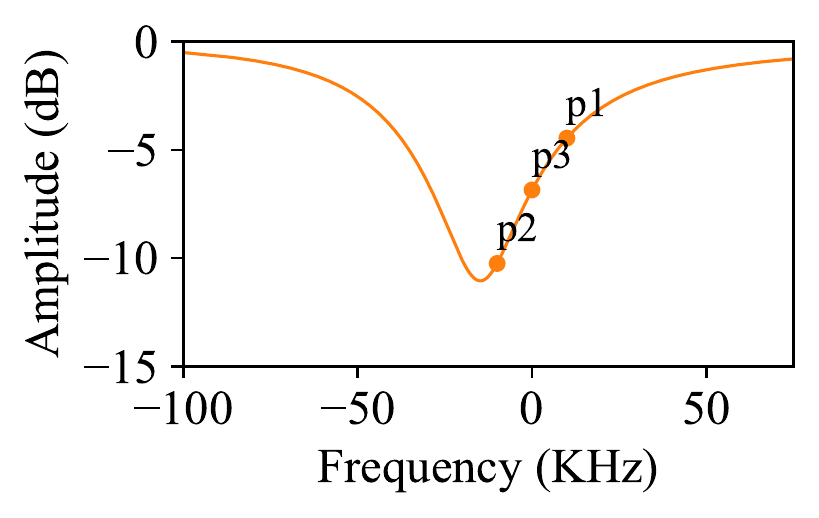}
    \vspace*{\baselineskip} 
    \includegraphics[width=0.68\linewidth]{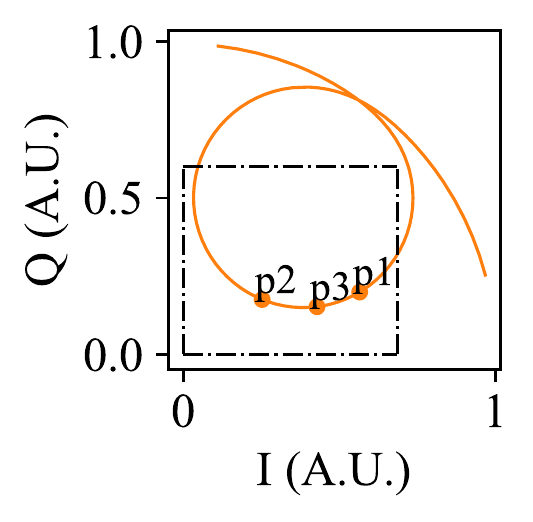}    
  \end{subfigure}
  \begin{subfigure}[b]{0.7\linewidth}
    \centering
    \includegraphics[width=0.89\linewidth]{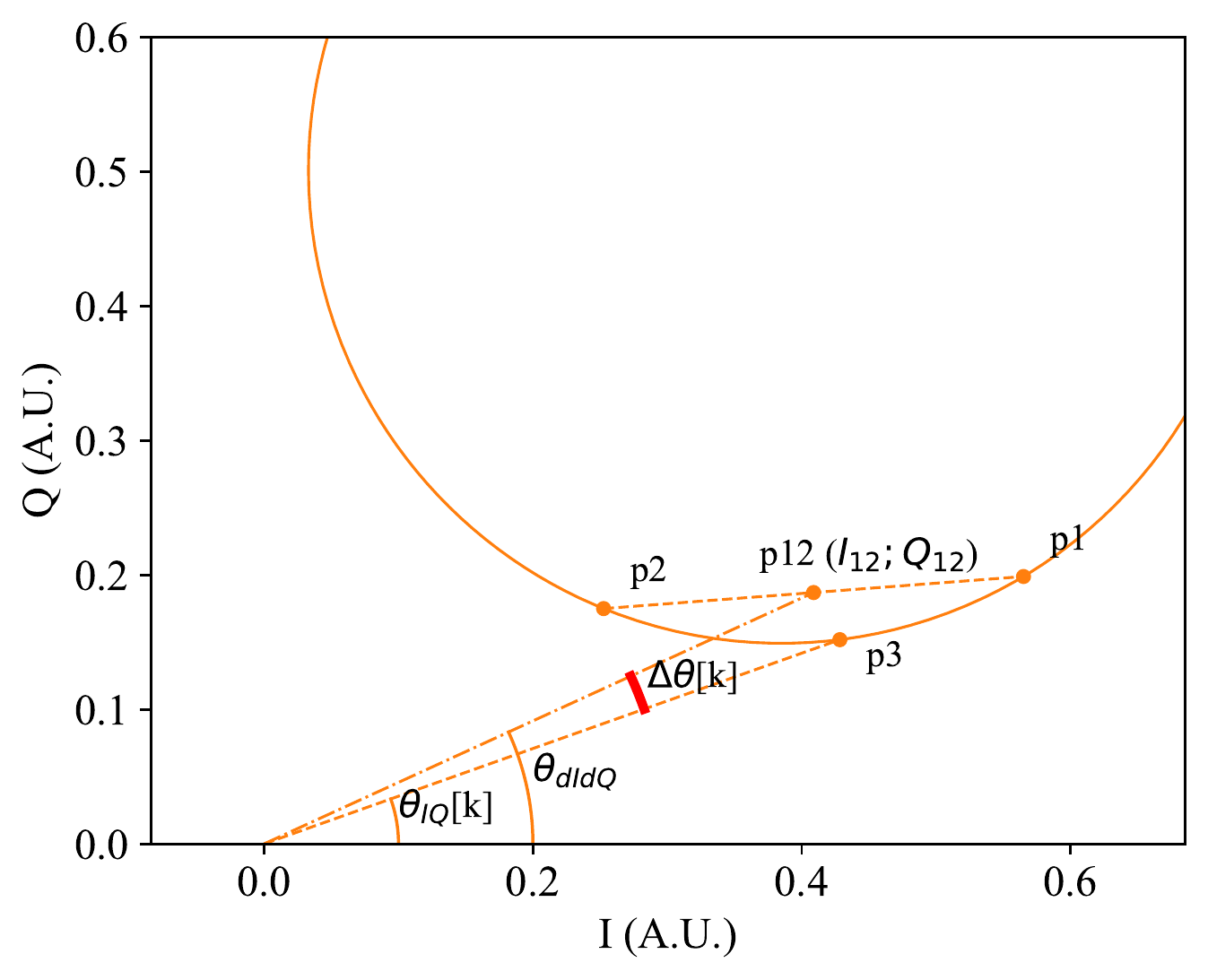}
  \end{subfigure}%
  \caption{The same but drifted LEKID, $p_3$ is away from the bottom of the resonance (F$_{tone}$ $\neq$ F$_{res}$). This is the case when $\Delta\theta$[k] $\neq$ 0.}
\label{figZoomROFF}
\end{figure}

The \shift\  procedure uses an algorithm fed by the data blocks featuring the three-point-modulation as discussed earlier.
For each LEKID, the difference in phase ($\Delta\theta$[k]) of the expected resonance center frequency that should lie at point $p_3$ is compared to the phase of the resulting median point $p_{12}$ computed with $p_1$ and $p_2$, see figures~\ref{figZoomRON} and \ref{figZoomROFF}.
\begin{equation} 
    \Delta\theta[k] = \theta_{IQ}[k] - \theta_{dIdQ} 
\label{eqDiffang}
\end{equation}
where $ \theta_{IQ}[k]=\arctan{\left( \frac{Q_3[k]}{I_3[k]} \right)}$ and
$\theta_{dIdQ}=\arctan{\left(\frac{Q_{12}}{I_{12}} \right)}$ with $(I_{12}, Q_{12}) = ( \frac{I_1[k] + I_2[k]}{2}, \frac{Q_1[k] + Q_2[k]}{2} )$.

\begin{figure}[hbtp] \centering
\begin{subfigure}[b]{0.32\textwidth} \centering
    \includegraphics[width=\textwidth]{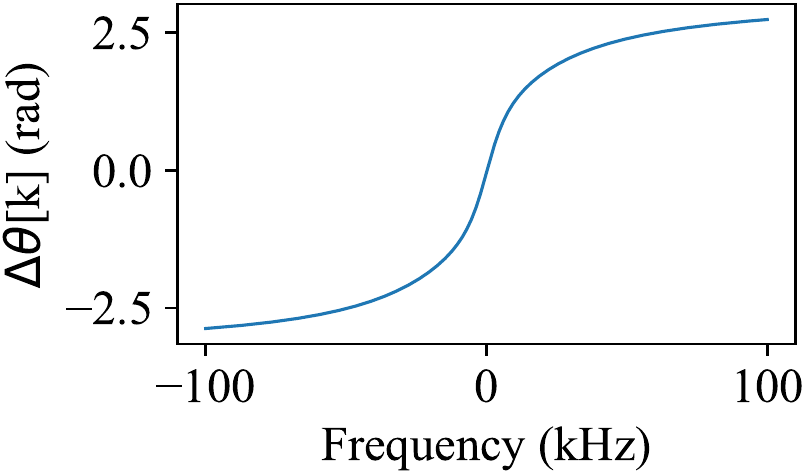} \caption{$\Delta\theta_k$ variation vs frequency.} \label{figTheta}
\end{subfigure}
\begin{subfigure}[b]{0.32\textwidth} \centering
    \includegraphics[width=\textwidth]{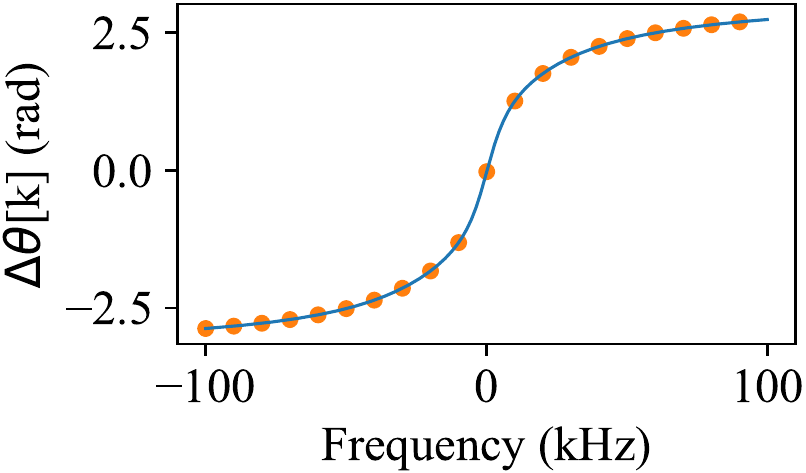} \caption{\locate\ with LO sweep.} \label{figLocate}
\end{subfigure}
\begin{subfigure}[b]{0.32\textwidth} \centering
    \includegraphics[width=\textwidth]{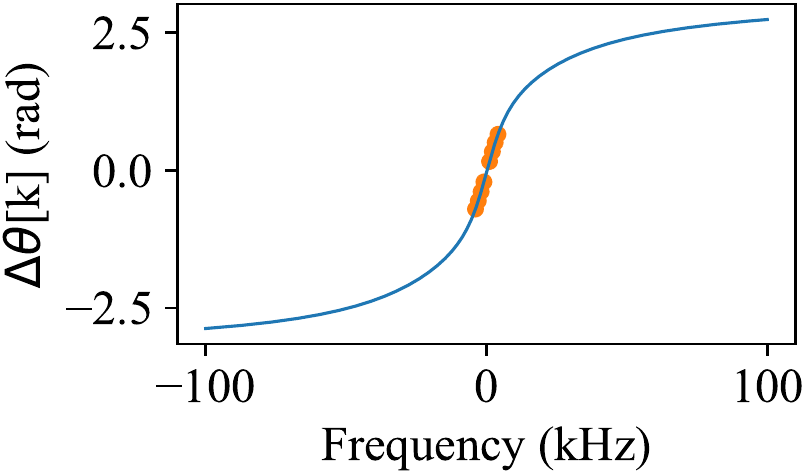} \caption{Real-time \follow\ } \label{figTonesFollow}
\end{subfigure}
\caption{{$\Delta\theta[k]$} variation over frequency and its use in the various \shift\  procedures.} \label{figThetaFollow}
\end{figure}

As depicted in figure~\ref{figTheta}, $\Delta\theta_k$ indicates how close the tone is to the LEKID resonance frequency such that:
\begin{itemize}
    \item if $\Delta\theta_k=0$ : The tone is well tuned
    \item if $\Delta\theta_k>0$ : F$_{tone}$ > F$_{res}$
    \item if $\Delta\theta_k<0$ : F$_{tone}$ < F$_{res}$
\end{itemize}
These quantities $\Delta\theta_k$ are then indirectly used to compute an average dephasing per feed-line $\Delta\theta_{avg}$ which is then used to operate frequency corrections on the electronic tones to make sure they, on average,match the LEKID resonances.

To  obtain a $\Delta\theta_{avg}$ per feed-line, we select the highest signal-to-noise LEKIDs, i.e. those
for which the magnitudes $M_{dIdQ}=\sqrt{I_{12}^2 + Q_{12}^2 }$ are the largest, as this is the signature of the deepest resonances. In our case we base the process on 10 LEKIDs ($N_{det}$=10) for each feedline as a tradeoff between precision and the processing capabilities.
For the $N_{det}$ selected LEKIDs, we use the $\Delta\theta_k$ values to calculate $Q_{avg}=\sum_{k=1}^{N_{det}} \cos{(\Delta\theta[k])}$ and $I_{avg}=\sum_{k=1}^{N_{det}} \sin{(\Delta\theta[k])}$ from which yields:
\begin{equation} \label{eqAngAvg}
    \Delta\theta_{avg} = \arctan{\left(\frac{Q_{avg}}{I_{avg}}\right)}
\end{equation}

Following the discussion above two \shift\ sub-modes named \locate\ and \follow\ are used during CONCERTO observations.

{\bf \locate\ } is used only at instrument startup. 
As shown in figure~\ref{figLocate}, it performs a limited 200\,kHz LO sweep with a fixed frequency step of 10\,kHz and for each of these LO steps, we compute $\Delta\theta_{avg}$.
Once the sweep is finished, we search the minimal $\Delta\theta_{avg}$ to find the optimal LO frequency value and from that we deduce the frequency shift to apply to the frequency comb.

{\bf \follow\ } is used between consecutive scans where the telescope may change orientation and thus the background change.
It ensures that the excitation frequency comb stays on LEKID resonances by quickly performing real-time global frequency adjustments.
In this mode, $\Delta\theta_{avg}$ is computed for each data block and the adequate frequency correction is applied, see figure~\ref{figTonesFollow}. 
That is when  $\Delta\theta_{avg}$ is zero no correction is applied, when positive the frequency is decreased and when negative the frequency is increased.
This global frequency correction is performed at roughly 2.4\,Hz with a maximal excursion of one digital frequency bin (3.8\,kHz).

\section{Individual LEKID'S resonance frequency tuning method: \tuning}

\begin{figure}[hbtp] \centering
    \includegraphics[ width=0.89\textwidth]{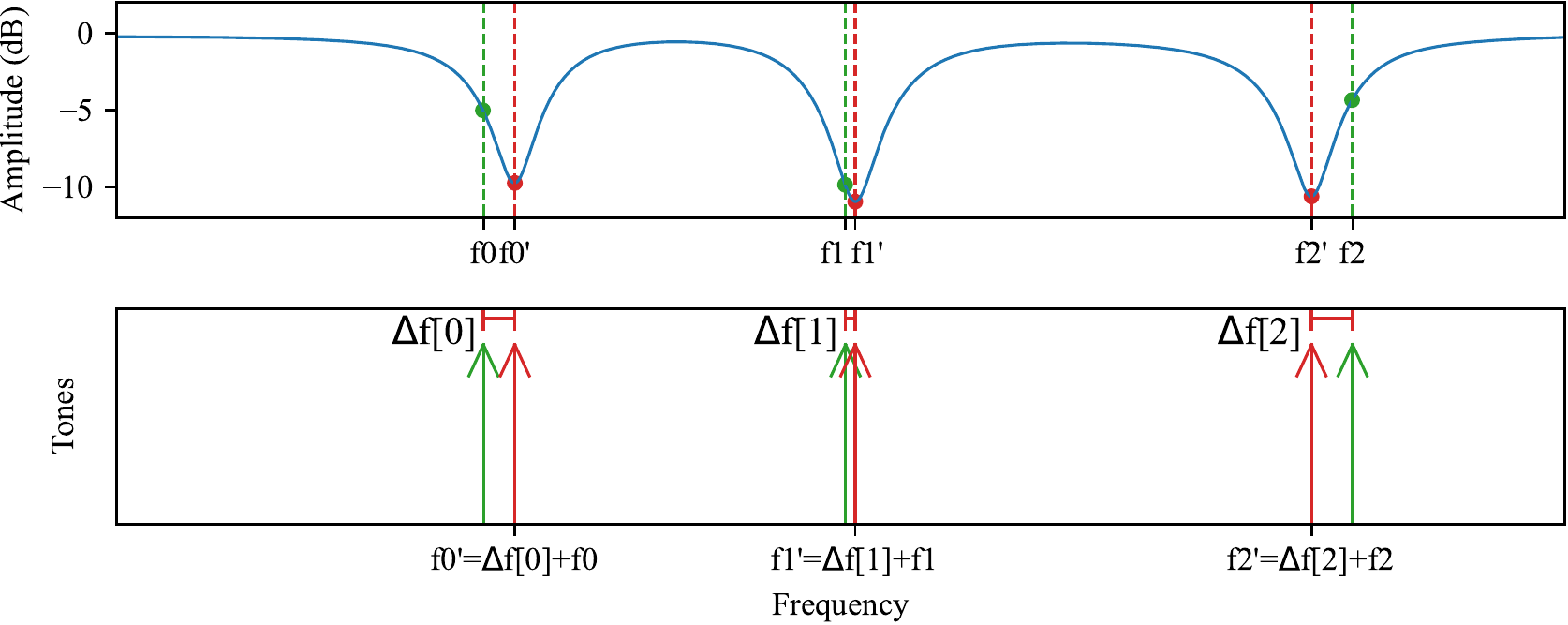}
\caption {The tones are corrected individually to match the resonance frequencies. In green and red we show the initial tone frequencies and those after applying the \tuning\ procedure. \label{figTuning}} 
\end{figure}

The main principle of astrophysical observations with the LEKIDs is to measure the resonance frequency shift induced by the incoming light from the targeted source. As a consequence, 
it is mandatory to set the tones precisely on each of the LEKIDs resonance frequencies just before starting a scan.  This can not be achieved neither with the \locate\ or \follow\ procedures, which implement a global shift.
Due to their extreme sensitivity, the LEKIDs do not react equally to background variations.
Thus, the so-called \tuning\ procedure
is used to correct the tones position individually as schematically shown in figure ~\ref{figTuning}. 
This correction is applied when the telescope has stopped moving
between consecutive scans, that is just when the new scan starts. 
Then, during observation data are taken with the frequency tones fixed to the values obtained from the \tuning. 

\begin{figure}[hbtp] \centering
    \includegraphics[ width=0.49\textwidth]{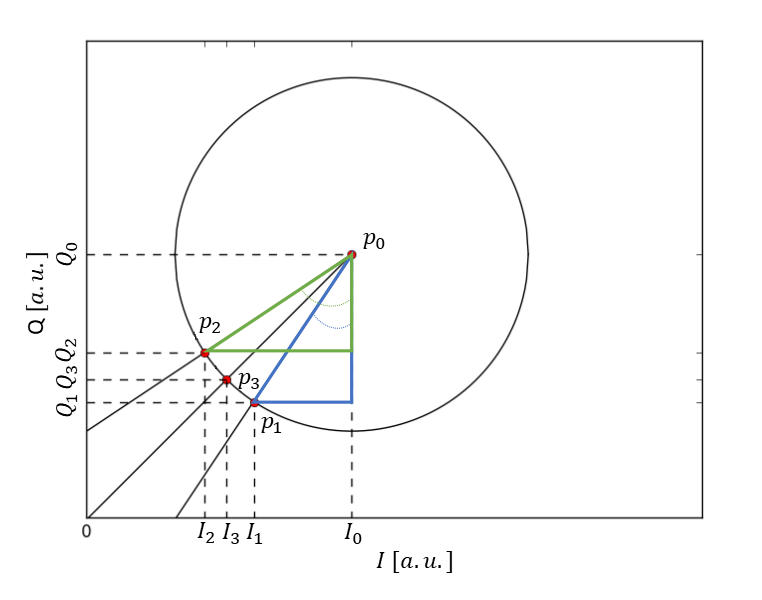}
\caption { $p_0$ ($I_0$;$Q_0$) is obtained by a circle fitting from $p_1$,$p_2$ and $p_3$ , adapted from \cite{fasano2021accurate} \label{figCircleFit}}
\end{figure}

\noindent The \tuning\ procedure is based on the calibration procedure presented in \cite{fasano2021accurate} and was also used for the KISS experiment \cite{Fasano_19JLTP_KISS}.
As illustrated in figure~\ref{figCircleFit} we first estimate a calibration coefficient, $C$, by fitting for each data block a circle to the mean modulation and scientific data signals ( $p_1=(I_1;Q_1)$, $p_2=(I_2;Q_2)$ and $p_3=(I_3;Q_3)$ in the figure):
\begin{equation}
		C =  \frac{2*F_{mod}}{\Delta\theta}
\end{equation}
\noindent From $\theta_{12} = \arctan\left( \frac{I_0-I_{12}}{Q_0 - Q_{12}}  \right)$ we define $\Delta \theta = \theta_2-\theta_1$ with $p_0=(I_0,Q_0)$ the center of the resonance circle.


\noindent From this we can now compute the difference in frequency between the current tone position for each detector and the expected frequency of resonance of the detector as:
\begin{equation}
\Delta f = \left( \theta_3 - \theta_0 \right) \times C
\end{equation}
where $\theta_0 = \arctan\left( \frac{I_0}{Q_0}  \right)$ and $\theta_3 = \arctan\left( \frac{I_0-I_3}{Q_0 - Q_3}  \right)$. \\

\noindent During the \tuning\ procedure the $\Delta f$ of each detector is converted into frequency bins and sent to the corresponding electronic board, which will adjust the input frequency tone for the detector accordingly.
To avoid propagating the noise in the estimation of $\Delta f$ and eventually placing the tone out of the LEKID's resonance frequency, we typically correct only for 70\,\% of the difference in frequency. Therefore we apply multiple corrections until it converges accurately.
Note that the \tuning\ procedure can only work if the frequency tones are placed within the resonance and relatively close to the frequency of resonance.
Furthermore, as discussed above it needs to be performed just before a scan to ensure an optimal starting point. Currently, we take about 30 seconds of \tuning\ before starting a new scan to successfully correct the tones.

\section{Conclusions}
LEKIDs are well adapted for multiplexing, minimizing the needs in terms of cold electronics, and so for astrophysical observations. In this regard, specific
developments of readout electronics and acquisition software are needed.
The major challenge to use LEKIDs in observational conditions 
is to be able to cope with background variations while ensuring a fast sampling rate (a few kHz).
We have developed a dedicated electronic readout and acquisition software to solve this issue. The solution is to use one single frequency tone per detector but implement the three-point-modulation technique. The latter allows us to instantaneously monitor the variations on the LEKID's resonance frequencies, and thus, adapt to the background variations.
For this, we use various procedures depending on both the instrumental and observation conditions. 
The first step is to evaluate the frequency of each tone of the readout by a \sweep\ to match the LEKIDs resonance frequencies, which is used as a reference. Furthermore, thanks to frequency modulation introduced by the LO, we are able to 
adjust on-line the tone frequencies to adapt to background variations during observations. 
We have implemented a fast algorithm, named \shift, for adapting to either global drifts with respect to the reference startup \locate, or to continuous variations as the telescope moves between scans \follow. Finally, to precisely and individually match the tones to the LEKIDs resonance frequencies
we have implemented a slower method, \tuning.
We conclude that the procedures presented in this paper allow us to successfully operate CONCERTO in the APEX telescope under standard observation conditions.

\section*{Acknowledgements}
We acknowledge support from the European Research Council (ERC) under the European Union’s Horizon 2020 research and innovation programme (project CONCERTO, grant agreement No 788212) and from the Excellence Initiative of Aix-Marseille University-A*Midex, a French “Investissements d’Avenir” programme.

\bibliography{Biblio}
\bibliographystyle{JHEP}

\end{document}